\documentclass[aps,prd,twocolumn,twoside,superscriptaddress,floatfix, nofootinbib]{revtex4-1}
\pdfoutput=1

\usepackage{amssymb, amsmath, bm, dcolumn, epsf, graphicx, latexsym, slashed, simplewick, ifthen}
\usepackage{float}
\usepackage[utf8]{inputenc}
\usepackage{subfigure}
\usepackage{comment}
\usepackage{graphicx}
\usepackage{hyperref}
\usepackage{enumerate}
\usepackage{color}
\usepackage{aas_macros}
\usepackage[usenames,dvipsnames]{xcolor}
\hypersetup{
    colorlinks = true,
    citecolor = {MidnightBlue},
    linkcolor = {BrickRed},
    urlcolor = {BrickRed}
}

\newcommand{\ba}{\begin{eqnarray}}
\newcommand{\ea}{\end{eqnarray}}

\newcommand{\beq} {\begin{equation}}
\newcommand{\eeq} {\end{equation}}
\newcommand{\bal} {\begin{aligned}}
\newcommand{\eal} {\end{aligned}}
\newcommand{\indep}{\perp \!\!\! \perp}
\newcommand{\lya}{{Lyman-$\alpha$} }

\begin{document}

\title{Two shadows on two backlights: 
forecasting the Ly-$\alpha$ forest $\times$ CMB Lensing bispectrum 
from ACT/SO/S4 \& DESI
}

\author{Adrien La Posta}
\affiliation{Department of Physics, University of Oxford, Denys Wilkinson Building, Keble Road, Oxford OX1 3RH, United Kingdom}
\affiliation{Université Paris-Saclay, CNRS/IN2P3, IJCLab, 91405 Orsay, France\\}

\author{Emmanuel Schaan}
\affiliation{Kavli Institute for Particle Astrophysics and Cosmology,
382 Via Pueblo Mall Stanford, CA 94305-4060, USA}
\affiliation{SLAC National Accelerator Laboratory 2575 Sand Hill Road Menlo Park, California 94025, USA}

%%%%%%%%%%%%%%%%%%%%%%%%%%%%%%%%%%%%%%%%%%%%%%%%%%%%%%%
\begin{abstract}
We forecast the sensitivity of future correlations between CMB lensing and the \lya forest power spectrum.
This squeezed-limit bispectrum probes the connection between the \lya transmission and the underlying large-scale matter density field.
We recover the measured signal-to-noise (SNR) ratio obtained with Planck$\times$BOSS, and forecast a SNR of $10.3, 14.8$ and $20.2$ for DESI combined with ACT, SO and CMB-S4 respectively.
For DESI and SO/CMB-S4, the correlation should be detectable at SNR$\ge 5$ in 5 redshift bins, useful for distinguishing our signal from several contaminants.
Indeed, our forecast is affected by large theoretical uncertainties in the current modeling of the signal and its contaminants, such as continuum misestimation bias or damped-\lya absorbers.
Quantifying these more accurately will be crucial to enable a reliable cosmological interpretation of this observable in future measurements.
We attempt to enumerate the features required in future \lya forest simulations required to do so.
\end{abstract}

\maketitle

%%%%%%%%%%%%%%%%%%%%%%%%%%%%%%%%%%%%%%%%%%%%%%%%%%%%%%%
\section{Introduction}

The bright light from distant quasars acts a backlight for the large-scale structure.
Fluctuations of the matter density field, accompanied by fluctuations in the neutral hydrogen density, imprint shadows on this backlight, in the form of the \lya absorption at a rest-frame wavelength of $121.6$~nm.
The redshifted \lya absorption produces a forest of lines in the quasar spectrum -- the so-called \lya forest.
Such ``skewers'' of the neutral hydrogen density are one of the best probes of the matter density fluctuations on the smallest scales.
As such, they have produced some of the tightest constraints on the masses of neutrinos and warm dark matter~\cite{viel2013,delabrouille2015,delabrouille2020,villasenor2023}
By cross correlating the \lya absorption across lines of sight, and with other tracers of the matter (e.g., quasars themselves), the \lya is also a powerful probe of the large-scale matter fluctuations at high redshift, including the baryonic acoustic oscillations.~\cite{busca2013,slosar2013,delubac2015,saintagathe2019,blomqvist2019,bourboux2020}

Crucially, all these cosmological results rely on understanding the connection between matter density and \lya absorption.
One generally assumes the neutral hydrogen causing the absorption to be in photoionization equilibrium in a uniform UV background, with no additional source of entropy in the hydrogen gas.
This leads to the fluctuating Gunn-Peterson approximation~\cite{croft1998}, routinely implemented in numerical simulations.~\cite{borde2014,rossi2014}

However, many complex physical effects are expected to affect the matter - \lya absorption relation. 
They include Jeans smoothing, thermal broadening and nonlinear peculiar velocities on small scales, the ionization from the proximity with UV sources (including the quasars) on large scales, and any source of entropy in the gas equation of state.~\cite{aip2015, Peeples10, garzilli2015}
Measuring empirically the matter - \lya connection would therefore be valuable.

To do this, we follow Refs.~\cite{valinotto2009,doux16} and consider a different imprint of the large-scale structure on a different backlight: the lensing distortion on the cosmic microwave background (CMB)~\cite{hu_okamoto}.
The CMB lensing convergence is a direct probe of the matter overdensity field, integrated along the line of sight.~\cite{lewis2006}
Cross-correlating it with the \lya forest is thus a direct way to measure their connection.~\cite{valinotto2009,vallinotto2011,doux16,chiang17}

Unfortunately, this direct cross-correlation is difficult in practice.
Uncertainty in the quasar continuum, i.e. the quasar spectrum we would see in the absence of any \lya absorption, is largely degenerate with the large-scale density Fourier modes along the line of sight.
Without these large-scale modes, one cannot estimate the mean absorption on a given \lya forest, which would most directly correlate with the CMB lensing convergence.
While this may be possible in the near future with new quasar continuum techniques, we here focus on a way to reconstruct the mean absorption along the \lya forest indirectly, via its nonlinear coupling to the small-scale modes we can measure in the \lya forest~\cite{chiang2014,lii2014,wagner2014}.
Physically, nonlinear evolution under gravity causes the small-scale \lya power spectrum to respond (linearly at first order) to the mean overdensity on the same line of sight. 
Estimating the $\kappa$-\lya correlation thus requires to face the same issues as for correlations between line intensity mapping and any integrated quantity such as CMB lensing or tSZ. 
Indeed, the large-scale modes are also suppressed in LIM due to galactic contaminants and most of the findings about the $\kappa$-\lya bispectrum should apply on LIM cross correlations.

In this paper, we revisit existing models for this cross-correlation and its contaminants, following Refs.~\cite{doux16,chiang18}, highlighting the modeling challenges and the ingredients which need to be simulated in order to accurately interpret the signal.
We perform a signal-to-noise estimation for current and upcoming experiments and find that DESI-Y5 and CMB lensing reconstructions from ACT, SO and CMB-S4 will lead to a significant detection of the signal in multiple redshift bins.

%%%%%%%%%%%%%%%%%%%%%%%%%%%%%%%%%%%%%%%%%%%%%%%%%%%%%%%
\section{Intensity maps with missing modes: enabling cross-correlations with 2D fields}

\begin{figure}[t!]
\centering
\includegraphics[width=0.8\columnwidth]{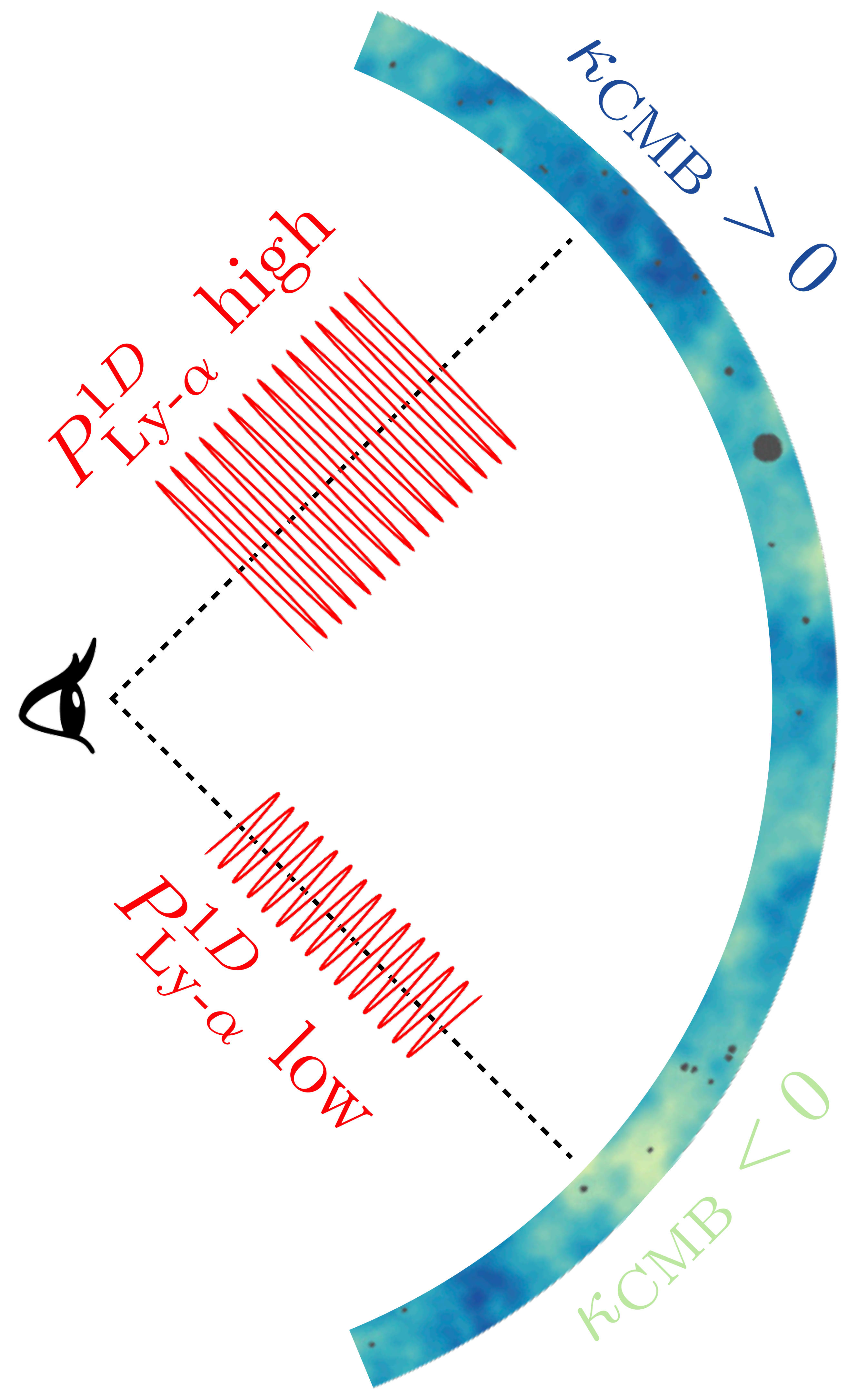}
\caption{
Because of the nonlinear evolution under gravity,
the amplitude of the \lya power spectrum on small scales responds to the large-scale matter overdensity.
We look for this signal by cross-correlating the CMB lensing convergence, probe of the matter overdensity, with the 1D \lya forest power spectrum.
This amounts to reconstructing the large-scale Fourier modes from the small-scale modes, with a quadratic estimator in the \lya forest, akin to tidal reconstruction~\cite{zhu2022}.}

\label{fig:schematic}
\end{figure}

To measure the \lya absorption within a given forest, one compares the observed, absorbed quasar spectrum with an estimate of the quasar continuum, i.e. the unabsorbed spectrum.
While the quasar continuum is uncertain, it can be assumed to be smooth.
As a result, the uncertain quasar continuum is degenerate with the large-scale line of sight modes of the one-dimensional \lya power spectrum.
These Fourier modes are thus typically discarded.
However, these are specifically the Fourier modes needed to measure the mean \lya absorption on the corresponding line of sight.
Without these modes, the \lya forest can thus not be correlated with two-dimensional tracers of the matter density such as CMB lensing.

This feature is generic for many line intensity maps, both in absorption (\lya forest) or emission (e.g., 21~cm, CO, CII, H$\alpha$ or \lya lines).
For emission line intensity maps, the issue if often continuum foregrounds, such as the zodiacal light, the Milky Way or the cosmic infrared background.
These foregrounds are typically assumed to be spectrally smooth, and thus only degenerate with the large-scale Fourier modes along the line of sight.
These modes are thus typically discarded, making it impossible to correlate these observables with 2D maps.
This is unfortunate, as such cross-correlations would be valuable in ascertaining the cosmological nature of observed line intensity mapping signals, and would constrain their connection to the matter density field.

This issue can be circumvented by better modeling the foregrounds, or the quasar continuum in the case of the \lya forest.
Another solution is to use the \lya forest or the intensity map as a light source whose lensing can be reconstructed.~\cite{schaan2018,abhishek2022}
This lensing can then be correlated with 2D maps.
Yet another solution is to reconstruct the missing line-of-sight modes, using their coupling to the small-scale Fourier modes which are directly observable.
This is generally done with a quadratic estimator, and called position-dependent power spectrum or tidal reconstruction.~\cite{chiang2014,doux16,zhu2022}

In this paper, we follow this approach.
We use the small-scale 1D \lya power spectrum as a quadratic estimator for the large-scale density field, and correlate it with CMB lensing.
This cross-correlation is thus effectively a bispectrum measurement.
We model this bispectrum following the position-dependent power spectrum formalism, as used in the derivation of the supersample variance.~\cite{takada2013,li2014,schaan2014}
\begin{equation}
    P_{\mathrm{Ly}\alpha}(k_\parallel, \mathbf{r_\perp}) = P_{\mathrm{Ly}\alpha}(k_\parallel) + \frac{\mathrm{d}P_{\mathrm{Ly}\alpha}(k_\parallel)}{\mathrm{d}\bar{\delta}}\bar{\delta}(\mathbf{r_\perp})
\end{equation}
where $\mathbf{r_\perp}$ is the angular position on the studied slice.
As a result,
\begin{equation}\label{eq:bispec}
    \langle P_{\mathrm{Ly}\alpha}(k_\parallel, \mathbf{k_\perp}) \kappa^*(\mathbf{k'_\perp}) \rangle = \frac{\mathrm{d}P_{\mathrm{Ly}\alpha}(k_\parallel)}{\mathrm{d}\bar{\delta}}P^{2\mathrm{D}}_{\bar{\delta}\kappa}(\mathbf{k_\perp})
\end{equation}
where $P^{2\mathrm{D}}_{\bar{\delta}\kappa}(\mathbf{k_\perp})$ is given by
\begin{equation}\label{eq:pdeltakappa}
    P^{2\mathrm{D}}_{\bar{\delta}\kappa}(\mathbf{k_\perp}) = \langle \kappa(\mathbf{k_\perp})\bar{\delta}(\mathbf{k_\perp}) \rangle.
\end{equation}

In order to evaluate these expressions for our signal of interest, we start with the CMB lensing convergence in Sec.~\ref{sec:cmb_lensing}.
We then focus on the \lya forest power spectrum in Sec.~\ref{sec:lya_boss_desi} and its response to a matter overdensity in \ref{sec:dplya_ddelta}, highlighting the important theoretical uncertainty in the expected signal.

%%%%%%%%%%%%%%%%%%%%%%%%%%%%%%%%%%%%%%%%%%%%%%%%%%%%%%%
\section{CMB lensing from Planck, ACT, SO \& CMB-S4}
\label{sec:cmb_lensing}

The signal of interest is a correlation of the one-dimensional \lya power spectrum with the lensing convergence measured in the same line of sight. 
We therefore seek an estimator for the lensing convergence on a given line of sight that minimizes its mean-squared error with respect to the true lensing convergence.
The estimator linear in the convergence map which minimizes this mean-squared error is the Wiener filter:
\begin{equation}
    \kappa_{\ell m}^\mathrm{WF} = W_\ell\kappa_{\ell m},
\end{equation}
where $W_\ell = C_\ell^{\kappa\kappa}/(C_\ell^{\kappa\kappa}+N_\ell^{\kappa\kappa})$. 

\begin{figure}[t!]
\centering
\includegraphics[width=\columnwidth]{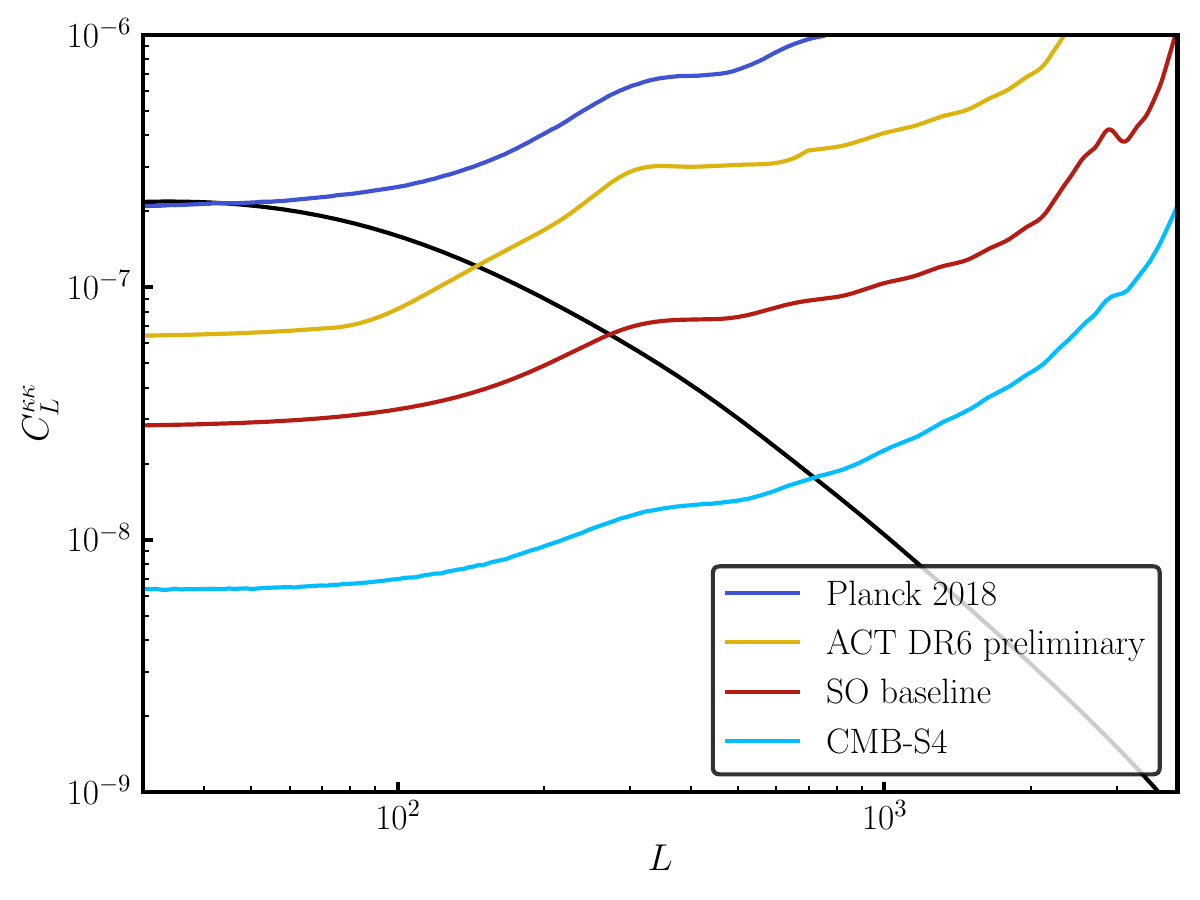}
\caption{
Planck, ACT, SO and CMB-S4 will measure the CMB lensing convergence power spectrum with ever lower noise power spectrum (colored curves).
The CMB lensing measurement is signal dominated per Fourier mode at every multipole $L$ where the noise power spectrum is below the expected lensing power spectrum (black curve, from a fiducial Planck 2018 TT+TE+EE cosmology).
}
\label{fig:lensing_noise_power}
\end{figure}

In principle, one could instead explores the scale dependence of the \lya-\lya-CMB lensing bispectrum, as a function of the separation between the separation between the position in the CMB lensing map and the \lya forest line of sight.
However, this scale dependence is likely trivial, given by the CMB lensing-matter correlation function, which is better probed directly with CMB lensing alone.
As a result, we choose to compress the information (and the signal-to-noise) from this scale-dependence into an optimal point estimate for the CMB lensing convergence, the Wiener-filtered convergence.

The resulting filtered map $\kappa^\mathrm{WF}(\hat{n})$ will therefore be a smoothed version of the original map with a typical smoothing scale $\theta_\mathrm{WF}$. 
We demonstrate in appendix~\ref{app:lensing_patches} that even if we have multiple 
\lya forest lines of sight separated by less than $\theta_\mathrm{WF}$ (i.e. with a highly correlated CMB lensing measurement) we do not lose constraining power on the $\kappa$-Ly$\alpha$ bispectrum. 
In the following we will show forecasted signals and errors for different dataset combinations. We will use the Planck 2018 lensing measurement to compare the result of our forecast with previous measurement on the data (using \lya forests from BOSS in~\cite{doux16}). 
We will present forecasts for combinations that use the recent ACT DR6 lensing measurements~\cite{qu2023,madhavacheril2023}, and the upcoming Simons Observatory~\cite{simons} and CMB-S4~\cite{cmbs4} measurements. 
The noise power spectra used in this analysis are shown in Fig.~\ref{fig:lensing_noise_power}.

\begin{figure}[t!]
\centering
\includegraphics[width=\columnwidth]{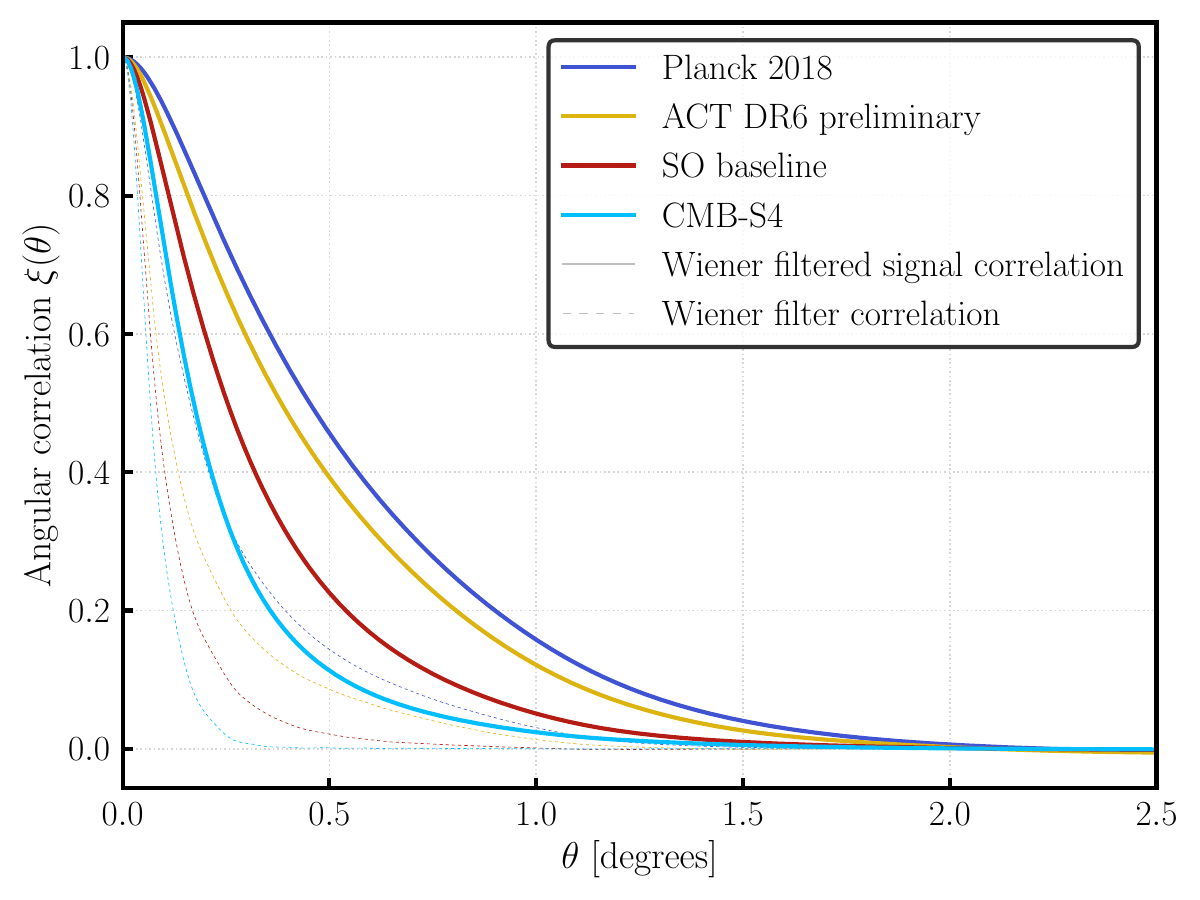}
\caption{We apply a Wiener filtering to the lensing convergence maps to maximize the local signal-to-noise ratio. 
In real space, the Wiener filtering is a smoothing of the map with a kernel (dashed lines). 
We then compute the angular correlation of the filtered signal (solid lines) for Planck 2018 (blue), ACT DR6 (yellow) and Simons Observatory (red).}
\label{fig:correlation_filtered_lensing}
\end{figure}

Figure~\ref{fig:correlation_filtered_lensing} shows the Wiener filter in real space as well as the correlation function of the Wiener filtered convergence maps. We choose to quantify the correlation length of the Wiener filtered $\kappa$ convergence map with $\xi(\theta_\mathrm{WF}) = 0.5$, which means that two directions separated by more than $\theta_\mathrm{WF}$ have less than $50\%$ correlation. 
We obtain $\theta_\mathrm{WF} = 0.53,\;0.44,\;0.27,\;0.18$ degrees for Planck, ACT DR6, SO and CMB-S4 respectively.
This scale is relevant, as it gives us an idea of the typical size of patches of roughly constant $\kappa^\text{WF}$.

Table~\ref{tab:varkappa} displays the standard deviation of the convergence map (i.e. the value of the correlation function for $\theta=0$). 
Note that the experiments with small noise power spectra have higher variance. 
Indeed the Wiener filter roughly set to $0$ the signal on noise dominated scales. 
The second column shows the squared SNR, ie the ratio of the variance coming from the signal, and the variance coming from the noise.
\begin{table}[b!]
\begin{tabular}{lcc}
\textbf{Experiment} & $\sigma(\hat{\kappa}_\mathrm{WF})$ & $\sigma_\mathrm{signal}/\sigma_\mathrm{noise}$\\
\hline
Planck & 0.0152 & 0.58\\
ACT DR6 & 0.0210 & 0.80\\
SO baseline & 0.0302 & 0.99\\
CMB-S4 & 0.0422 & 1.41\\
\end{tabular}
\caption{
Standard deviation of the Wiener-filtered convergence field. As the CMB lensing sensitivity improves with Planck, ACT, SO and CMB-S4,
fewer and fewer scales are filtered out by the Wiener filter,
thus increasing the ratio of signal to noise (right column)
and increasing the overall variance (middle column).
}
\label{tab:varkappa}
\end{table}

We can express the lensing convergence $\kappa$ in Eq.~\ref{eq:pdeltakappa} as the integrated matter density perturbation, such that the two-dimensional power spectrum becomes
\begin{equation}
    P^{2\mathrm{D}}_{\bar{\delta}\kappa}(\mathbf{k_\perp}) = \Lambda_\kappa(-\mathbf{k_\perp})\int_{\chi_\mathrm{min}}^{\chi_\mathrm{max}}d\chi ' W_\kappa(\chi ') F(\chi,\chi',\mathbf{k_\perp}),
    \label{eq:p2dlens}
\end{equation}
where
\begin{equation}
    F(\chi,\chi',\mathbf{k_\perp}) = \int \frac{dq}{2\pi} P_{\delta\delta}(q, \mathbf{k_\perp})\mathrm{sinc}(\frac{q\Delta\chi}{2})e^{iq(\chi' - \chi)}.
\end{equation}
The quality of the CMB lensing reconstruction appears only as the Wiener filter in Fourier space $\Lambda_\kappa(-\mathbf{k_\perp})$ in Eq.~\ref{eq:p2dlens} (see Appendix~\ref{app:model} for more details.)
It implies that the higher the lensing signal-to-noise is, the higher the bispectrum signal (Eq.~\ref{eq:bispec}) between CMB lensing and the \lya power spectrum  will be.

%%%%%%%%%%%%%%%%%%%%%%%%%%%%%%%%%%%%%%%%%%%%%%%%%%%%%%%
\section{\lya forest from BOSS \& DESI}
\label{sec:lya_boss_desi}

\begin{figure}[t!]
\centering
\includegraphics[width=\columnwidth]{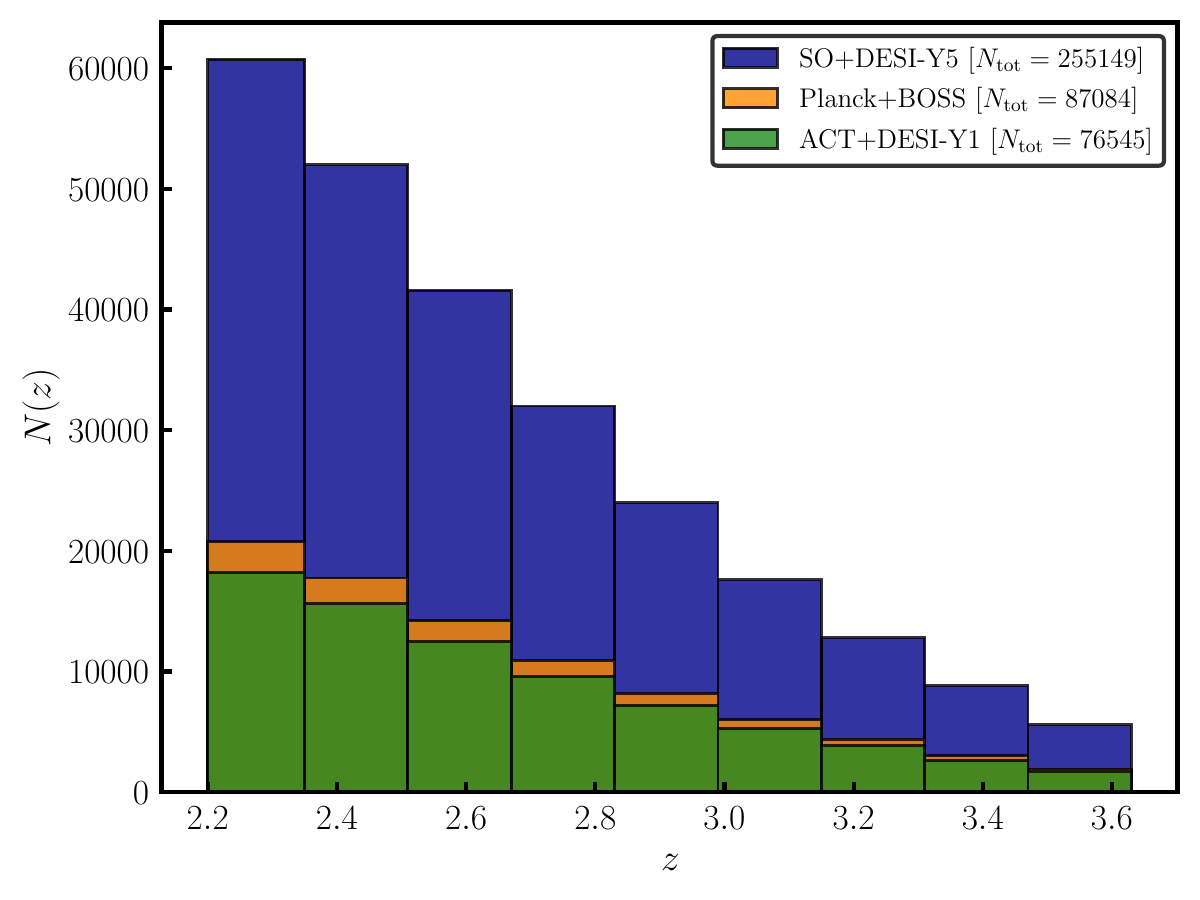}
\caption{Number of galaxies per redshift bin for DESI~\footnote{obtained from the public github repository \url{https://github.com/julienguy/simplelyaforecast}}. If we integrate the number density, it corresponds to $51$ lines of sight per square degree. 
We show here the total number of galaxies per redshift bin considering different survey areas. 
To get the correct number density for our Planck+BOSS forecast we rescaled this density to get the same total number of lines of sight than in Ref.~\cite{doux16}.}
\label{fig:number_density}
\end{figure}

The one-dimensional \lya power spectrum $P_{\mathrm{Ly}\alpha}^{1D}$ does not explicitly enter the expression for the $\kappa-\mathrm{Ly}\alpha$ bispectrum as shown in Eq.~\ref{eq:bispec}. 
However, it directly enters the covariance of the bispectrum.
This can be approximated as:
\begin{equation}
    \mathrm{cov}(B_{\kappa,\mathrm{Ly}\alpha}) = \mathrm{cov}(P_{\mathrm{Ly}\alpha}^{1D}) \times \mathrm{cov}(\hat{\kappa}_\mathrm{WF}),
\label{eq:cov}
\end{equation}
which depends on the variance of the $\kappa$ convergence field and the variance of the one-dimensional $\mathrm{Ly}\alpha$ power spectrum. 
This expression holds when the two quantities $\kappa$ and $P_{\mathrm{Ly}\alpha}$ are approximately independent.
This is verified by computing their correlation coefficient for an individual line of sight.
This correlation coefficient is effectively the SNR per line of sight for our effect, which is very small.

We thus average a large number line of sights to improve the signal-to-noise on the bispectrum. 
Forecasting this thus requires knowing the number of l.o.s per redshift bin. 
Fig.~\ref{fig:number_density} displays the total number of l.o.s for the 9 redshift bins considered in this analysis. We assumed a survey area of $1500$ square degrees for the ACT+DESI-Y1 forecast and $5000$ square degrees for the SO+DESI-Y5 forecast. 
The total number density of the assumed quasars redshift distribution is $51$ l.o.s. per square degree. 
The number density for BOSS is computed assuming that the total number of forests is $87,085$ as quoted in~\cite{doux16} to enable a fair comparison with that measurement of the bispectrum.

Different lines of sight with small angular separation will be multiplied by highly correlated values of $\kappa^\text{WF}$ in our bispectrum estimator.
One may worry that the total SNR would therefore not scale as $\sqrt{N_\mathrm{los}}$, but instead as the shallower $\sqrt{N_{\kappa^\text{WF}}}$, where $N_{\kappa^\text{WF}}$ is the number of patches with independent values of $N_{\kappa^\text{WF}}$.
In Appendix~\ref{app:covariance_matrix} we demonstrate that this is not the case and that we indeed expect the signal-to-noise on the bispectrum to improve with $\sqrt{N_\mathrm{los}}$ and not with the number of independent $\kappa$ patches.
\begin{figure}[t!]
\centering
\includegraphics[width=\columnwidth]{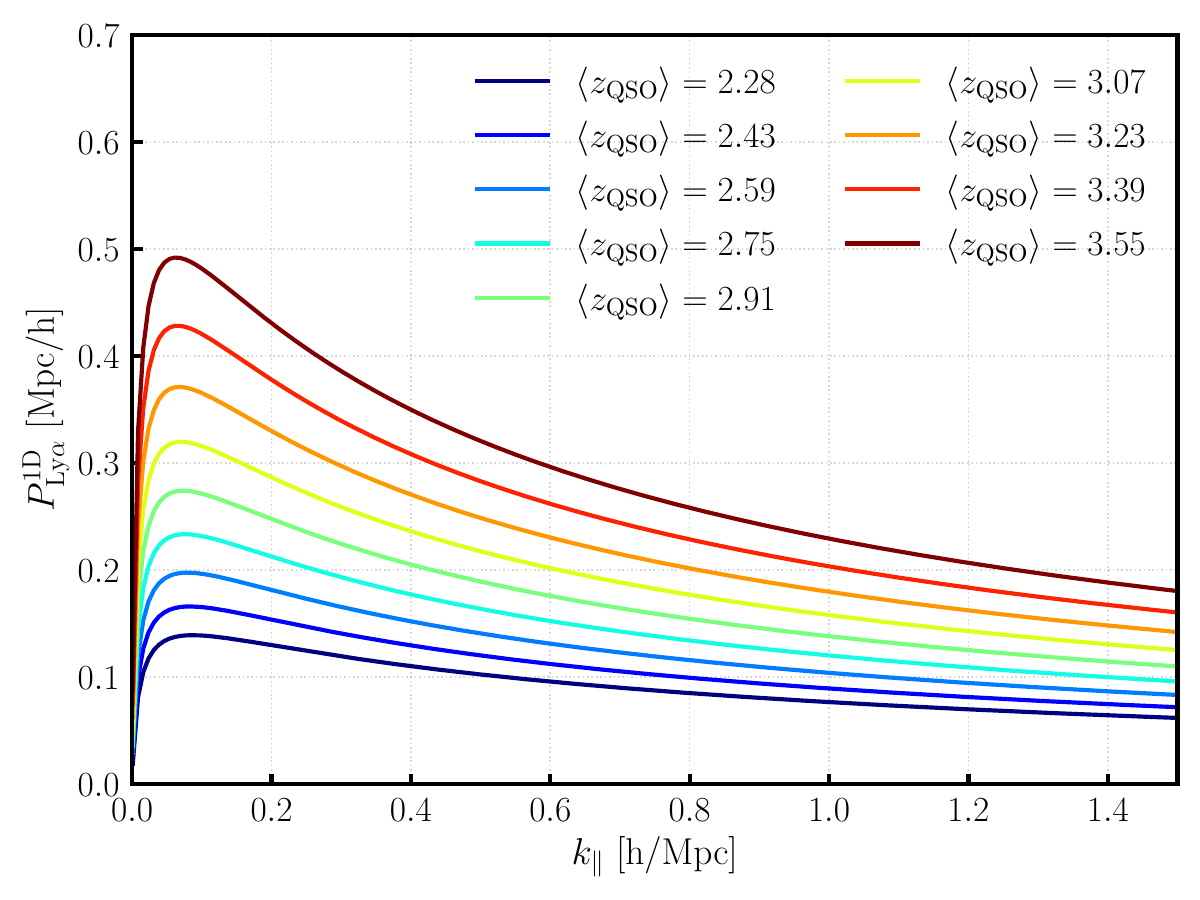}
\caption{In order to forecast the uncertainty on the $\kappa - \mathrm{Ly}\alpha$ bispectrum, we need to estimate the uncertainty on the one-dimensional \lya power spectrum. 
We choose the NPD13 model~\cite{delabrouille13} for the 1D power spectra, shown here for the bin center used in this forecast.}
\label{fig:p1d_lya}
\end{figure}

The variance of the one-dimension power spectrum is given by
\begin{equation}
    \mathrm{cov}(P_{\mathrm{Ly}\alpha}^{1D}) = \frac{2(P_{\mathrm{Ly}\alpha}^{1D}+P_\mathrm{noise}/W_\mathrm{spectro}^2)^2}{N_\mathrm{modes}},
\end{equation}
where $N_\mathrm{modes} = \Delta k * \Delta\chi / (2\pi)$ is the number of modes per $k$ bins and 
$W_\mathrm{spectro} \equiv \text{exp}\left( - k_\parallel^2 \sigma_R^2 /2 \right) \text{sinc}\left( k_\parallel \delta \chi/2 \right)$
is the spectrograph window function accounting for resolution effects. 
Throughout the paper, we use a spectrograph resolution $\sigma_R=0.77 \;\mathrm{Mpc}/h$ estimated from Ref.~\cite{delabrouille13} for BOSS data, and a typical pixel size $\delta \chi = 2\pi/k_{\parallel \text{max}}$ with 
$k_{\parallel \text{max}} = 6 h/$Mpc. 
We account for DESI's higher spectral resolution ($50\%$ improvement with respect to BOSS~\cite{desi_instr_ov}), by rescaling the spectrograph resolution quoted above.
Our baseline model for the one-dimensional power spectrum is the NPD13 template~\cite{delabrouille13}. 
The model power spectra are shown in Fig.~\ref{fig:p1d_lya} for different quasar redshifts.

\begin{figure}[t!]
\centering
\includegraphics[width=\columnwidth]{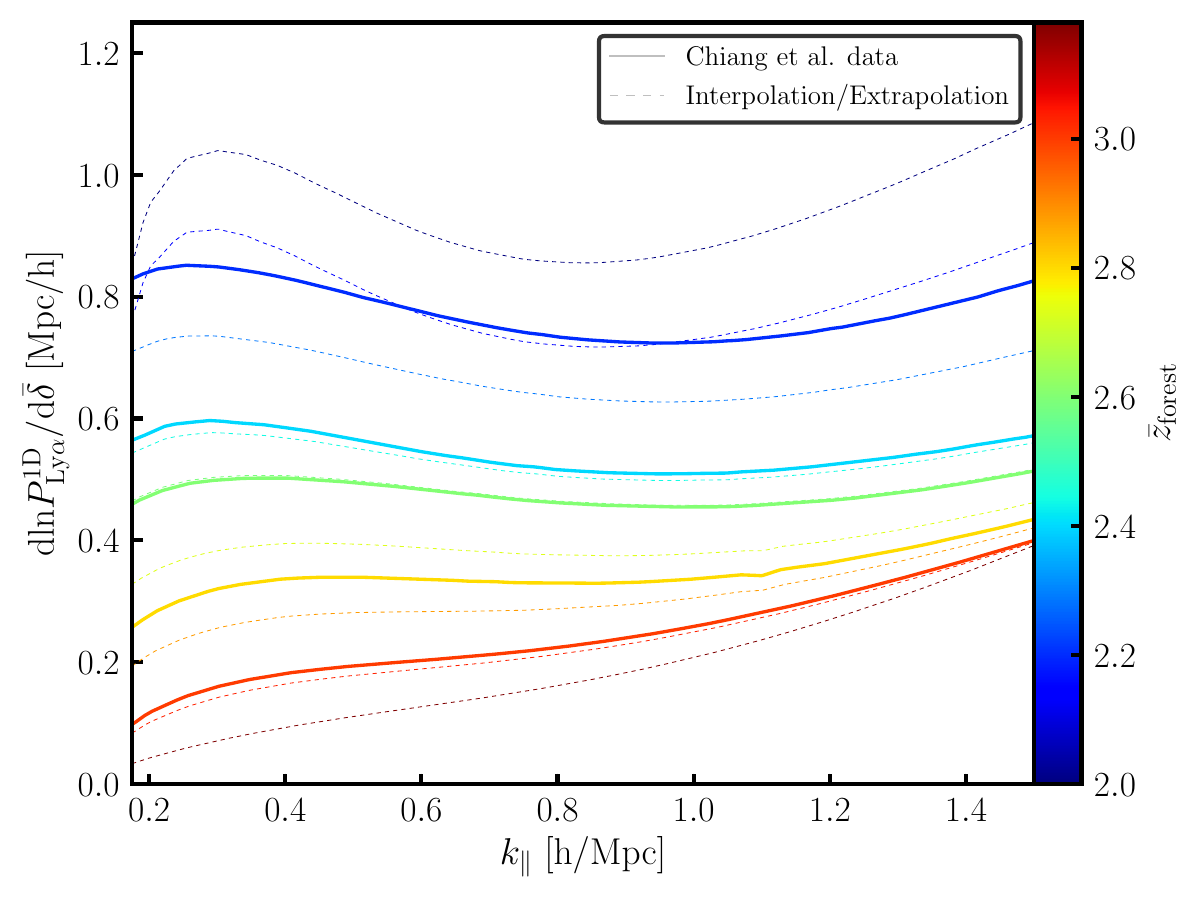}
\caption{
        Logarithmic response of the \lya one-dimensional power spectrum as measured in Ref.~\cite{chiang18} from separate universe simulation as described in Ref.~\cite{chiang17,cieplak2016}. The response measured on simulations in C18 are displayed in bold lines for $z=2.2$, $2.4$, $2.6$, $2.8$ and $3.0$. For the purpose of this forecast, we interpolate the response in the range $2.2 \le z \le 3.0$ and extrapolate it outside using the fitting formulae Eq.~\ref{eq:a_of_z} and Eq.~\ref{eq:b_of_z}. The thin lines correspond to our interpolation/extrapolation of the response for redshifts $z=2.01$, $2.15$, $2.29$,$ 2.44$, $2.59$, $2.73$, $2.88$, $3.03$ and $3.18$. One should note that the redshifts discussed here correspond to the center of the forest $\bar{z}_\mathrm{forest}$.
}
\label{fig:dlogp1d_chiang17}
\end{figure}

The noise power spectrum is given by

\begin{equation}
    P_\mathrm{noise}(k, z) = \frac{c}{(\mathrm{SNR}_\mathrm{QSO})^2\lambda_{\mathrm{Ly}\alpha}H(z)},
\end{equation}
where $\mathrm{SNR}_\mathrm{QSO}$ is the wavelength-averaged SNR of individual quasar spectra in unit of $1/\sqrt{\text{\r{A}}}$. 
The mean SNR used in this forecast, averaging over all redshift bins is~$3.0\;{\text{\r{A}}}^{-0.5}$. We find it to be in agreement with SNR values quoted in the literature for DESI~\cite{ravoux_lya_1d, desi_lya_snr_sim}. 
We expect the spectrograph noise to mostly matter at low-$z$ and at small scales along the line-of-sight. Indeed we have checked that rescaling the $\mathrm{SNR}_\mathrm{QSO}^2$ by a factor of $2$ induces a $30\%$ variation of the forecasted bispecrum SNR for the lowest redshift bin, and only of order $10\%$ for the highest redshift bins.
We perform average within the redshift bins considered in the analysis to get the noise power spectrum used to compute the variance of the one-dimensional power spectrum in that bin.

%%%%%%%%%%%%%%%%%%%%%%%%%%%%%%%%%%%%%%%%%%%%%%%%%%%%%%%
\section{Response of the \lya power spectrum to a local overdensity}
\label{sec:dplya_ddelta}

The response 
$dP_{\mathrm{Ly}\alpha}^\mathrm{1D}/d\bar{\delta}$
of the one-dimensional \lya power spectrum to a local matter overdensity is needed to evaluate the bispectrum as expressed in Eq.~\ref{eq:bispec}. 
As we show in this section, there are large theoretical uncertainties on the modeling of this quantity.
These uncertainties propagate to uncertainties in our forecasted SNR, and will limit our ability to interpret future measurements.
Narrowing down the theoretical uncertainty, e.g., with simulations~\cite{chiang18}, will therefore be crucial.
In this section, we outline the theoretical ingredients needed and their uncertainties.
%
%%%%%%%%%%%%%%%%%%%%%%%%%%%%%%%%%%%%%%%%%%%%%%%%%%%%%%%
\subsection{Baseline model: separate-Universe simulations from Chiang et al.}

Our baseline model for the 
$P_{\mathrm{Ly}\alpha}^\mathrm{1D}$
response to a large-scale overdensity comes from Ref.~\cite{chiang18}.
It uses the separate Universe approach~\cite{chiang17,cieplak2016} to estimate the signal, shown in Fig.~2 of Ref.~\cite{chiang18}.

Figure~\ref{fig:dlogp1d_chiang17} displays $R(k, z) = d\mathrm{log}P^\mathrm{1D}/d\bar{\delta}$ evaluated for forests centered on different redshifts. 
Ref.~\cite{chiang18} provides simulations of this quantity at $z=2.2$, $2.4$, $2.6$, $2.8$ and $3.0$ displayed as thick lines. 
To estimate the response at an arbitrary redshift, we use bilinear interpolation for $2.2 \le z \le 3.0$. 
To compute the response outside of this redshift range, we have extrapolated the quantity $R(k,z)/R(k, z_\mathrm{ref})$ where $z_\mathrm{ref}$ is a reference redshift (set to $z=3$ in the following). 
The extrapolation model is defined as $\frac{R(k,z)}{R(k,z_\mathrm{ref})} = k^{a(z)}e^{b(z)}$ where
\begin{equation}
    \label{eq:a_of_z}
    a(z) = a_1 \left( \frac{1+z}{1+z_\mathrm{ref}} - 1 \right) \left( \frac{1+z}{1+z_\mathrm{ref}} \right)^{a_2},
\end{equation}
and
\begin{equation}
    \label{eq:b_of_z}
    b(z) = b_1 \left( \frac{1+z}{1+z_\mathrm{ref}} - 1 \right).
\end{equation}
The interpolated and extrapolated quantities are shown in Fig.~\ref{fig:dlogp1d_chiang17} as thin dashed lines. 
We choose to interpolate the logarithmic response of the one-dimensional Ly$\alpha$ power spectum 
$d\mathrm{ln}P/d\bar\delta = dP/d\bar\delta / P$
rather than
$dP/d\bar\delta$
to be less affected by potential differences in the absolute amplitude of the power spectrum $P$ between the simulations and the model that we are using in this paper~\cite{delabrouille13}. 
The arbitrary choices made during extrapolation do not impact significantly the forecasted signal-to-noise presented in this paper because the signal-to-noise in the highest redshift bin ($z \geq 3$) is small (see Section~\ref{sec:results}).

The simulations on which this model is based model a number of complex physical effects.
These include Jeans smoothing where baryonic pressure support erases fluctuations on small scales,
thermal broadening of the \lya absorption lines due to local thermal velocity dispersion,
redshift-space distortions on large scales (Kaiser effect) and small scales (Fingers of God),
and finally the nonlinear structure formation under gravity.
Many of these effects require an accurate modeling of baryonic physics, including feedback from galaxy formation, which are highly uncertain.
Furthermore, simulations with complex and realistic UV background fluctuations would be valuable in assessing how much these additional sources of fluctuations in the neutral hydrogen density could be detectable in our cross-correlation with CMB lensing.
In order to improve the modelling of the signal, ideally one would measure the squeezed bispectrum $B_{\kappa\text{\lya}}$ from simulations with a range of baryonic models over the scales of interest ($0.1 \le k_\parallel \le 1.5\;h/\mathrm{Mpc}$) and redshifts of interest ($2 \le z \le 3.6$) for DESI. 
Short of this, separate universe simulations with varying barynionic models would be useful to get a better estimate of the response of the one-dimensional \lya power spectrum to a large-scale overdensity, over the same range of scales and redshifts. Furthermore, several contaminating signals will have to be disentangled from our signal, in order to accurately interpret it. We discuss these below.

%%%%%%%%%%%%%%%%%%%%%%%%%%%%%%%%%%%%%%%%%%%%%%%%%%%%%%%
\subsection{Contaminant 1: Continuum fitting}

If the continuum fitting error was only an additive term, it would only affect the large scales in the \lya forest, which we are already discarding. 
Instead, the continuum fitting also enters in the normalization of the \lya forest on all scales, including the small scales we consider here. 

Indeed, as derived in appendix~\ref{app:cont_misestim}, the continuum variation across quasars is perfectly degenerate with the mean flux transmission along each quasar. Moreover, this mean flux transmission normalizes the fractional transmission fluctuation $\delta_F$.
This causes an additional term in the response of the power spectrum to an overdensity, which we estimate following Ref.~\cite{chiang18}. 
A misestimation of the continuum affects the observed flux perturbation $\hat{\delta}_F$ such that 
$\hat{\delta}_F = \delta_F (1-\Delta_F)$, 
where 
$\delta_F$ 
is the true \lya flux perturbation, and 
$\Delta_F$ 
is the mean flux transmission along the given \lya forest, lost due to the continuum misestimation.
This produces the following first-order correction to the measured power spectrum
\begin{align}
    \hat{P}_{FF}(k_\parallel, \mathbf{r}_\perp) &= \langle \hat{\delta}_F\hat{\delta}_F \rangle \nonumber \\
    &= P_{FF}(k_\parallel)\left[ 1 - 2\Delta_F(\mathbf{r}_\perp, \Delta\chi) \right].
\end{align}
Since the mean flux transmission $\Delta_F$ correlates with the matter overdensity, and hence with the CMB lensing convergence,
this leads to a bias in the $\kappa$-Ly$\alpha$ bispectrum~\cite{chiang18}:
\begin{align}
    \langle &\hat{P}_{\mathrm{Ly}\alpha}(k_\parallel)\kappa(\mathbf{r}_\perp) \rangle_\text{bias} \nonumber\\
    &= 
    - 2P_{\mathrm{Ly}\alpha}(k_\parallel)\int \frac{d^2\mathbf{k}_\perp}{(2\pi)^2} e^{-i\mathbf{k}_\perp\mathbf{r}_\perp} P^{2\mathrm{D}}_{\Delta_F\kappa}(\mathbf{k_\perp}).
\end{align}

%%%%%%%%%%%%%%%%%%%%%%%%%%%%%%%%%%%%%%%%%%%%%%%%%%%%%%%
\subsection{Contaminant 2: Damped \lya systems}

High column density \lya absorbers are among the main astrophysical contaminants to the \lya signal. 
Light from distant quasars may be absorbed by dense neutral hydrogen clouds on the line of sight. 
High column density absorbers are defined given a neutral hydrogen $H\mathrm{I}$ density threshold of $N_{H_\mathrm{I}} \ge 1.6\times 10^{17} \;\mathrm{atom}.\mathrm{cm}^{-2}$. 
A particular class of high density absorbers are the damped \lya absorbers (DLAs) with $N_{H_\mathrm{I}} \ge 2\times 10^{20}\;\mathrm{atom}.\mathrm{cm}^{-2}$ (see Ref.~\cite{wolfe05} for a review).
These not only saturate the \lya absorption at the wavelength corresponding to their redshifts, but they also affect nearby wavelengths due to their damping wings. If not accounted for, absorption from the damping wings of these absorbers will be incorrectly attributed to the presence of neutral hydrogen at these neighboring redshifts.

In overdense region, we expect the number of high column density absorbers to be larger than in underdense line of sights. 
This effect will correlate to the $\kappa$ convergence field which probes matter density along the line of sight, and therefore be a nuisance for the measurement of the $\kappa$-\lya bispectrum.
As discussed in Appendix~\ref{app:dla_effect}, adapted from Ref.~\cite{chiang18}, the effect of the high column density absorbers on the $\kappa$-\lya bispectrum may be modeled as
\begin{equation}
    \Delta B_{\kappa\mathrm{Ly}\alpha}^\mathrm{DLA}(k_\parallel) = b_\mathrm{DLA} P_{1\mathrm{D}}^\mathrm{DLA}(k_\parallel) \sigma^2_{\kappa\bar{\delta}}
\end{equation}
where $P_{1\mathrm{D}}^\mathrm{DLA}$ is the contribution of residual DLAs (including sub-DLAs and Lyman limit systems (LLS)) to the one-dimensional power spectrum.
We follow the prescription of Ref.~\cite{chiang18} and consider only Lyman limit systems (LLS) with $1.6\times 10^{17} \le N_{H_\mathrm{I}} \le 1\times 10^{19}\;\mathrm{atoms}.\mathrm{cm}^{-2}$ and sub-DLAs with $1\times 10^{19} \le N_{H_\mathrm{I}} \le 2\times 10^{20}\;\mathrm{atoms}.\mathrm{cm}^{-2}$.
This assumption holds as long as the higher density absorbers have been correctly identified and masked. To estimate the DLA contribution to the one-dimensional power spectrum, we use the fitting functions from Ref.~\cite{rogers17} and $b_\mathrm{DLA} = 2.17$ from Ref.~\cite{fontribeira2012} as done in Ref.~\cite{chiang18}.
\begin{figure}[H]
\centering
\includegraphics[width=\columnwidth]{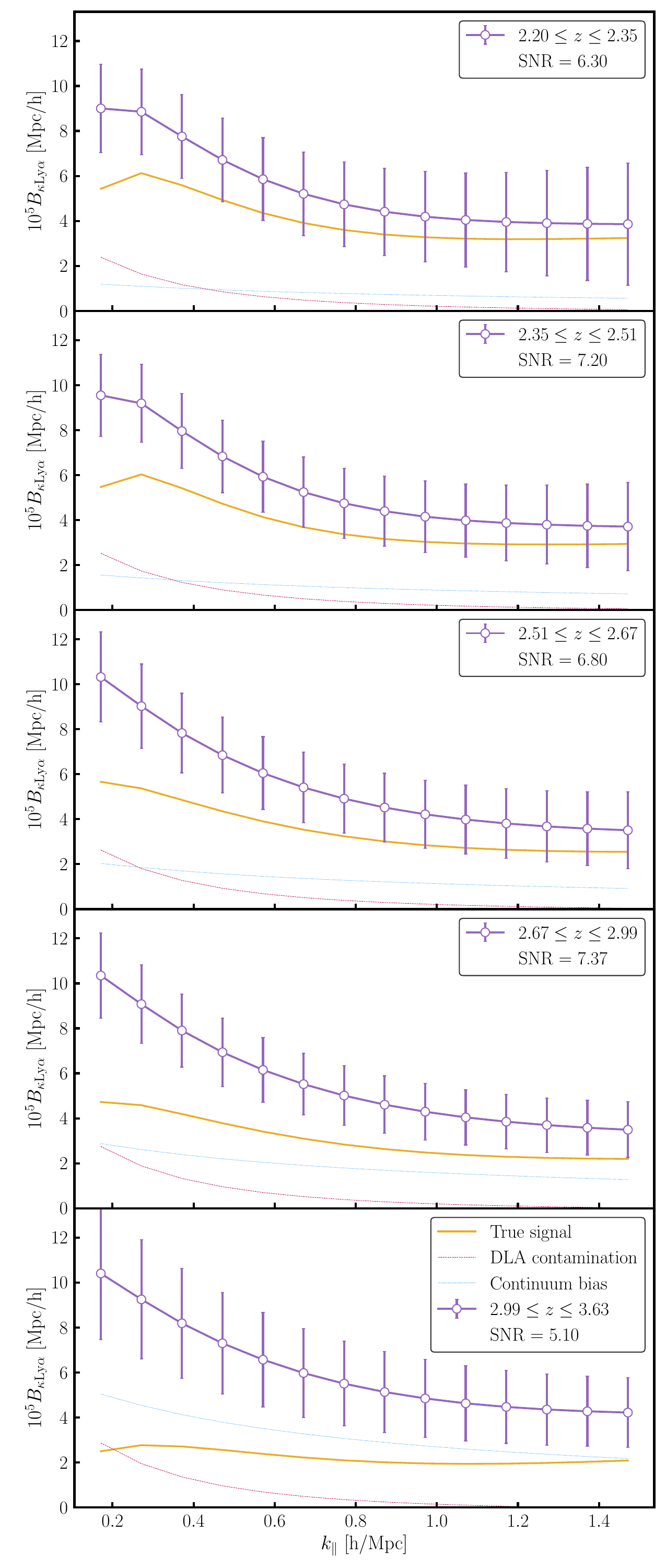}
\caption{Forecasted $\kappa$-$\mathrm{Ly}\alpha$ bispectrum from the combination Simons Observatory lensing measurements with DESI-Y5 $\mathrm{Ly}\alpha$ forests.
The signal-to-noise ratio is visibly sufficient to detect the correlation in five redshift bins.
However, the contaminating signals are comparable to the signal of interest, or even larger at the higher redshifts.
They will need to be accurately modelled and subtracted.
}
\label{fig:bispectrum_forecast_so_desiy5}
\end{figure}

%%%%%%%%%%%%%%%%%%%%%%%%%%%%%%%%%%%%%%%%%%%%%%%%%%%%%%%
\section{Forecast results}\label{sec:results}

\begin{table}[hbp!]
\begin{ruledtabular}
\begin{tabular}{lcccccc}
 &\multicolumn{6}{c}{\textbf{SNR}}\\
\textbf{Dataset} & $z_1$ & $z_2$ & $z_3$ & $z_4$ & $z_5$ & Combined\\
\hline
Planck+BOSS & 1.9 & 2.2 & 2.0 & 2.2 & 1.5 & \textbf{4.4}\\
ACT+DESI-Y1 & 2.4 & 2.8 & 2.6 & 2.8 & 1.9 & \textbf{5.7}\\
ACT+DESI-Y5 & 4.5 & 5.1 & 4.8 & 5.1 & 3.5 &  \textbf{10.3}\\
SO+DESI-Y5 & 6.3 & 7.2 & 6.8 & 7.4 & 5.1 & \textbf{14.8}\\
CMB-S4+DESI-Y5 & 8.5 & 9.8 & 9.3 & 10.2 & 7.1 & \textbf{20.2} \\
\end{tabular}
\end{ruledtabular}
\caption{Signal-to-noise ratios in different redshift bins. We display the SNR computed using a covariance matrix with a structure following what is shown in Doux+15. We use several redshift bins $z_1$ ($2.20 \le z \le 2.35$), $z_2$ ($2.35 \le z \le 2.51$), $z_3$ ($2.51 \le z \le 2.67$), $z_4$ ($2.67 \le z \le 2.99$), $z_5$ ($2.99 \le z \le 3.63$) and also show the signal-to-noise for the coadded bispectrum.
\label{tab:snr_binned}
}
\end{table}

In Section~\ref{sec:dplya_ddelta} we have described our method to estimate the response of the one-dimensional \lya power spectrum to a large scale matter overdensity. 
We have calculated the uncertainty on $P_{\mathrm{Ly}\alpha}^\mathrm{1D}$ in Section~\ref{sec:lya_boss_desi} and on the lensing convergence variance on the line of sight in Section~\ref{sec:cmb_lensing}. 
We are now in a position to compute the forecasted SNR given a lensing noise power spectrum and a survey area. 
In Fig.~\ref{fig:bispectrum_forecast_so_desiy5} we display the forecasted bispectrum along with the associated error bars as a purple line, for SO and DESI. 
The underlying cosmological signal is displayed as a solid yellow line and the contribution of contaminants is displayed as dashed and dotted lines. 
One can note that the contribution of the cosmological signal to the total bispectrum decreases with redshift. 
As depicted in Fig.~\ref{fig:bispectrum_forecast_so_desiy5}, the different contributions to the total bispectrum do not have the same redshift dependence. 
Therefore, it will be necessary to measure this signal in several redshift bins to be able to disentangle the underlying cosmological signal and the contaminants in order to provide the best constraints on the $P^\mathrm{1D}$ response to a large-scale matter overdensity.
This forecast shows that upcoming cosmological data from SO and DESI will enable such a measurement of the $\kappa$-$\mathrm{Ly}\alpha$ bispectrum with high significance ($\mathrm{SNR}>5)$ in several redshift bins.

Our expression Eq.~\eqref{eq:cov} for the covariance of the bispectrum does not assume any correlation between $k$-bins. 
In order to compute the signal-to-noise ratios presented in Table~\ref{tab:snr_binned}, we have assumed a correlation structure interpolated from Ref.~\cite{doux16}. 
We explore variation from this assumption in Appendix~\ref{app:correlations_snr}, defining three different scenarios : no correlation (i.e. a diagonal covariance matrix), a constant $30\%$ correlation and a structured correlation matrix inspired from Ref.~\cite{doux16}.

%%%%%%%%%%%%%%%%%%%%%%%%%%%%%%%%%%%%%%%%%%%%%%%%%%%%%%%
\section{Conclusions}

The cross-correlation signal between the \lya forest absorption and CMB lensing has been detected in Ref.~\cite{doux16} and may be a powerful probe to test the connection between the \lya transmission and the matter density field.
This is crucial for applications of the \lya forest which aim at constraining neutrino masses or deviations to the $\Lambda$CDM model such as warm dark matter.

Another motivation for developing estimators and understanding this bispectrum, is to use the \lya forest as a sandbox for the correlations of future intensity maps with projected quantities such as CMB lensing. 
Indeed, the problem of missing the large-scale Fourier modes along the line of sight is very generic to line intensity mapping and prevents from correlating line intensity maps with 2D projected maps like CMB lensing or thermal Sunyaev Zel'dovich (SZ). 
For the \lya forest, these modes are lost because they are degenerate with the continuum spectrum of the quasars. 
In 21cm intensity mapping, the large-scale modes are lost because they are degenerate with Galactic foregrounds. 
Thus, the method considered here, which effectively reconstructs the large-scale modes from quadratic combinations of the small-scale modes, should have applications to perform cross-correlations of CMB lensing or SZ with most of line intensity maps. 
Investigating the complexity and the limitations in the case of the \lya forest is a useful first step towards enabling line intensity mapping $\times$ CMB lensing correlations.

The goal of this study was to forecast the SNR for current and upcoming surveys, and to highlight the various sources of uncertainties in the theoretical modeling of this observable, so as to serve as a guide for future modeling efforts. 
Our forecast finds that the signal can be measured at a high signal-to-noise ratio ($SNR \ge 5$) in five redshift bins using DESI-Y5 \lya forests and the Simons Observatory CMB lensing. 
Despite the large uncertainties in the modeling of the signal itself, our forecasted analysis recovers the signal-to-noise ratio (SNR) obtained in the existing Planck$\times$BOSS measurement \cite{doux16} within $15\%$ ($4.4\sigma$ forecasted VS $5.0\sigma$ measured).

As mentioned above, our analysis highlights the large current uncertainty in the modeling of the signal itself as well as its contaminants. 
First, we would need a proper modeling of the impact of the linear and non-linear clustering bias as well as the impact of RSD and their response to a large-scale matter overdensity. 
For the purpose of this work, we relied on simulations to quantify these effects. 
To get an accurate modeling of the $\kappa$-\lya bispectrum signal measured from DESI \lya forests, one would need to have a set of separate universe simulations (with scales spanning the range $0.1 \le k \le 1.5 \;h/\mathrm{Mpc}$) for $2.2 \le z \le 3.6$ or simulations from which the bispectrum with CMB lensing can be estimated directly (though it may require a larger simulation box). 
The resolution along the line of sight should match the DESI resolution. 
As discussed above, some contaminants, such as continuum misestimation or DLAs, are comparable in size to the signal of interest. 
Thus having a proper modeling of these effects would be a useful addition to the simulation. 
Finally, it may be interesting to have different implementations of baryonic effects, including the feedback from galaxy formation, to study their impact on the $\kappa$-\lya bispectrum measurement and see if this observable is able to discriminate between them.

%%%%%%%%%%%%%%%%%%%%%%%%%%%%%%%%%%%%%%%%%%%%%%%%%%%%%%%
\section*{Acknowledgments}

We thank Roger de Belsunce, Sol\`ene Chabanier, Cyrille Doux, Simone Ferraro, Satya Gontcho-A-Gontcho, Julien Guy, Vid Irsic, Thibaut Louis, Frank Qu, Blake Sherwin, Tan Ting.
This work received support from the U.S. Department of Energy under contract number DE-AC02-76SF00515 to SLAC National Accelerator Laboratory.

%%%%%%%%%%%%%%%%%%%%%%%%%%%%%%%%%%%%%%%%%%%%%%%%%%%%%%%

\bibliographystyle{prsty.bst}
\bibliography{refs}
%%%%%%%%%%%%%%%%%%%%%%%%%%%%%%%%%%%%%%%%%%%%%%%%%%%%%%%

\onecolumngrid

\appendix

%%%%%%%%%%%%%%%%%%%%%%%%%%%%%%%%%%%%%%%%%%%%%%%%%%%%%%%
\newpage
\section{Modelling the signal and biases}
\label{app:model}

The purpose of this appendix is to give more details about the modelling of the bispectrum between the Lyman-$\alpha$ forest and CMB lensing $B_{\kappa,\mathrm{Ly}\alpha} = \langle P_{\mathrm{Ly}\alpha}\kappa\rangle$.

\subsection{Variation of the forest power spectrum}
The one-dimensional Lyman-$\alpha$ forest power spectrum will be modulated by large scale density fluctuations. We model this effect with the position-dependent power spectrum

\begin{equation}\label{eq:plya_modulation}
    P_{\mathrm{Ly}\alpha}(k_\parallel, \mathbf{r_\perp}) = P_{\mathrm{Ly}\alpha}(k_\parallel) + \frac{\mathrm{d}P_{\mathrm{Ly}\alpha}(k_\parallel)}{\mathrm{d}\bar{\delta}}\bar{\delta}(\mathbf{r_\perp}),
\end{equation}
where $\mathbf{r_\perp}$ is the angular position on the studied slice and $\bar{\delta}(\mathbf{r_\perp})$ is the 2D projected large scale density fluctuation, expressed as the average of the underlying density field along the line of sight

\begin{equation}\label{eq:deltabar}
    \bar{\delta}(\mathbf{r_\perp}) = \frac{1}{\Delta\chi}\int_{\chi_\mathrm{min}}^{\chi_\mathrm{max}} d\chi'\delta(\chi', \mathbf{r_\perp}).
\end{equation}

\subsection{Lensing in the studied slice}

The lensing convergence $\kappa$, can be described as a weighted average of the matter density field, introducing the lensing kernel $W(\chi)$ such that $\kappa(\mathbf{r_\perp}) = \int d\chi W(\chi)\delta(\chi, \mathbf{r_\perp})$. To maximize the signal to noise of the CMB lensing signal, we apply a Wiener filter on the lensing $\kappa$ maps such that $\hat{\kappa} = \Lambda_\kappa * \kappa + n$. Accounting for this operation, the 2D projected CMB lensing convergence is therefore

\begin{equation}\label{eq:kappa_wf_def}
    \kappa(\mathbf{r_\perp}) = \int_{\chi_\mathrm{min}}^{\chi_\mathrm{max}} d\chi' \int \mathrm{d}^2\mathbf{r'_\perp} \delta(\mathbf{r'})W_\kappa(\chi')\Lambda_\kappa(\mathbf{r'_\perp}-\mathbf{r_\perp})
\end{equation}\\   

\subsection{3-pt correlation function between $P_{\mathrm{Ly}\alpha}$ and $\kappa$}

Given Eq.~\ref{eq:plya_modulation}, we can derive an expression for the bispectrum in Fourier space in the transverse direction

\begin{equation}
    \langle P_{\mathrm{Ly}\alpha}(k_\parallel, \mathbf{k_\perp}) \kappa^*(\mathbf{k'_\perp}) \rangle = \frac{\mathrm{d}P_{\mathrm{Ly}\alpha}(k_\parallel)}{\mathrm{d}\bar{\delta}}P^{2\mathrm{D}}_{\bar{\delta}\kappa}(\mathbf{k_\perp})
\end{equation}
where $P^{2\mathrm{D}}_{\bar{\delta}\kappa}(\mathbf{k_\perp})$ is given by
    
\begin{align}
    P^{2\mathrm{D}}_{\bar{\delta}\kappa}(\mathbf{k_\perp}) &= \langle \bar{\delta}(\mathbf{k_\perp})\kappa^*(\mathbf{k_\perp}) \rangle \nonumber\\
    &= \Lambda(-\mathbf{k_\perp}) \int_{\chi - \frac{\Delta\chi}{2}}^{\chi + \frac{\Delta\chi}{2}} d\chi' W(\chi')\int\frac{dq}{2\pi} P_\mathrm{lin}(q, \mathbf{k_\perp})\mathrm{sinc}\left( \frac{q\Delta\chi}{2} \right) e^{iq(\chi' - \chi)},
\end{align}
which can be derived from Eq.~\ref{eq:deltabar} and Eq.~\ref{eq:kappa_wf_def}.\\

Finally, the $\kappa$-$\mathrm{Ly}\alpha$ bispectrum is given by

\begin{equation}
    \langle P_{\mathrm{Ly}\alpha}(k_\parallel)\kappa \rangle = \frac{\mathrm{d}P_{\mathrm{Ly}\alpha}(k_\parallel)}{\mathrm{d}\bar{\delta}} \int \frac{d^2\mathbf{k}_\perp}{(2\pi)^2} P^{2\mathrm{D}}_{\bar{\delta}\kappa}(\mathbf{k_\perp}).
\end{equation}

\begin{align}
    P_{\bar{\delta}\kappa}(\mathbf{k_\perp}) = \Lambda(-\mathbf{k_\perp}) \int_{\chi - \frac{\Delta\chi}{2}}^{\chi + \frac{\Delta\chi}{2}} d\chi_2 W(\chi_2)\int\frac{dq}{2\pi} P_\mathrm{lin}(q, \mathbf{k_\perp})\mathrm{sinc}\left( \frac{q\Delta\chi}{2} \right) e^{iq(\chi_2 - \chi)}
\end{align}

\subsection{Continuum misestimation bias}\label{app:cont_misestim}
As done in Ref.~\cite{chiang18}, the observed Ly-$\alpha$ flux can be modeled as

\begin{equation}
    F(\chi, \mathbf{r}_\perp) = A(\mathbf{r}_\perp)\bar{C}\bar{F}(r_\perp)(1 + \delta_F(\chi, \mathbf{r}_\perp))
\end{equation}
where $A$ is the brightness of the quasar, $\bar{C}$ is the mean continuum, $\bar{F}$ is the true mean flux, and $\delta_F$ is the true fluctuation around $\bar{F}$.\\

An estimator of $A$ can be written as

\begin{equation}
    \hat{A}(\mathbf{r}_\perp) = \frac{1}{\Delta\chi} \int_{\chi - \Delta\chi/2}^{\chi+\Delta\chi/2} d\chi' \frac{F(\chi', \mathbf{r}_\perp)}{\bar{C}(\chi')\bar{F}(\chi')} = A(\mathbf{r}_\perp)\left[ 1 + \frac{1}{\Delta\chi}\int_{\chi - \Delta\chi/2}^{\chi+\Delta\chi/2} d\chi' \delta_F(\chi', \mathbf{r}_\perp) \right],
\end{equation}
from which we can derive an estimator for $\delta_F$

\begin{equation}
    \hat{\delta}_F(\chi', \mathbf{r}_\perp) = \frac{F(\chi', \mathbf{r}_\perp)}{\hat{A}(\mathbf{r}_\perp)\bar{C}(\chi')\bar{F}(\chi')} - 1 = \frac{1 + \delta_F(\chi', \mathbf{r}_\perp)}{1 + \Delta_F(\mathbf{r}_\perp, \Delta\chi)} - 1 = \delta_F(\chi', \mathbf{r}_\perp) \left[ 1 - \Delta_F(\mathbf{r}_\perp, \Delta\chi) \right] - \Delta_F(\mathbf{r}_\perp, \Delta\chi).
\end{equation}

The Ly-$\alpha$ forest power spectrum is then

\begin{equation}
    \hat{P}_{FF}(k_\parallel, \mathbf{r}_\perp) = \langle \hat{\delta}_F\hat{\delta}_F \rangle = P_{FF}(k_\parallel)\left[ 1 - 2\Delta_F(\mathbf{r}_\perp, \Delta\chi) \right],
\end{equation}
Such that the biased measurement of the bispectrum is 
\begin{equation}
    \langle \hat{P}_{\mathrm{Ly}\alpha}(k_\parallel)\kappa(\mathbf{r}_\perp) \rangle = \frac{\mathrm{d}P_{\mathrm{Ly}\alpha}(k_\parallel)}{\mathrm{d}\bar{\delta}} \int \frac{d^2\mathbf{k}_\perp}{(2\pi)^2} e^{-i\mathbf{k}_\perp\mathbf{r}_\perp} P^{2\mathrm{D}}_{\bar{\delta}\kappa}(\mathbf{k_\perp}) - 2P_{\mathrm{Ly}\alpha}(k_\parallel)\int \frac{d^2\mathbf{k}_\perp}{(2\pi)^2} e^{-i\mathbf{k}_\perp\mathbf{r}_\perp} P^{2\mathrm{D}}_{\Delta_F\kappa}(\mathbf{k_\perp}).
\end{equation}

\subsection{DLA bias}\label{app:dla_effect}

In Ref.~\cite{rogers17}, the effect of DLA contamination on the 1D power spectrum is modelled as
\begin{equation}
    \frac{P_\mathrm{tot}^\mathrm{1D}}{P_{\mathrm{Ly}\alpha}} = \alpha_F + \sum_{i \neq F} \alpha_i(z)\frac{P_i}{P_{\mathrm{Ly}\alpha}}
\end{equation}
where 
\begin{equation}
    \frac{P_i}{P_{\mathrm{Ly}\alpha}} = \left( \frac{1 + z}{1 + z_0} \right)^{-3.55} \frac{1}{(a(z)e^{b(z)k_\parallel} - 1)^2} + c(z)
\end{equation}
In Ref.~\cite{chiang18}, they only consider contributions from \emph{subDLA} and \emph{LLS} as major contributions have already been removed in the data. Therefore, the total $\mathrm{Ly}\alpha$ power spectrum is 
\begin{equation}
    \frac{P_{\mathrm{Ly}\alpha}^\mathrm{biased}}{P_{\mathrm{Ly}\alpha}^\mathrm{unbiased}} = \alpha_F + \alpha_\mathrm{subDLA}\frac{P_\mathrm{subDLA}}{P_{\mathrm{Ly}\alpha}^\mathrm{unbiased}} + \alpha_\mathrm{LLS}\frac{P_\mathrm{LLS}}{P_{\mathrm{Ly}\alpha}^\mathrm{unbiased}}
\end{equation}
The bias introduced by DLAs in the $\mathrm{Ly}\alpha$ power spectrum can be expressed as

\begin{align}
    \Delta P_{\mathrm{Ly}\alpha}^\mathrm{DLA} &= P_{\mathrm{Ly}\alpha}^\mathrm{biased} - P_{\mathrm{Ly}\alpha}^\mathrm{unbiased} = \frac{P_{\mathrm{Ly}\alpha}^\mathrm{biased}}{P_{\mathrm{Ly}\alpha}^\mathrm{unbiased}} P_{\mathrm{Ly}\alpha}^\mathrm{unbiased} - P_{\mathrm{Ly}\alpha}^\mathrm{unbiased}\\
    &= P_{\mathrm{Ly}\alpha}^\mathrm{unbiased} \left[ \frac{P_{\mathrm{Ly}\alpha}^\mathrm{biased}}{P_{\mathrm{Ly}\alpha}^\mathrm{unbiased}} - 1 \right]\\
    &= P_{\mathrm{Ly}\alpha}^\mathrm{unbiased} \left[ \alpha_F - 1 + \alpha_\mathrm{subDLA}\frac{P_\mathrm{subDLA}}{P_{\mathrm{Ly}\alpha}^\mathrm{unbiased}} + \alpha_\mathrm{LLS}\frac{P_\mathrm{LLS}}{P_{\mathrm{Ly}\alpha}^\mathrm{unbiased}}  \right].
\end{align}

This bias will affect the one-dimensional power spectrum as
\begin{equation}
    \tilde{P}^{\mathrm{1D}}_{\mathrm{Ly}\alpha}(k_\parallel, \mathbf{r}_\perp) = P^{\mathrm{1D}}_{\mathrm{Ly}\alpha}(k_\parallel, \mathbf{r}_\perp) + \left(1 + b_\mathrm{DLA}\bar{\delta}(\mathbf{r}_\perp)\right) \Delta P_{\mathrm{Ly}\alpha}^\mathrm{DLA}(k_\parallel),
\end{equation}
resulting in an additional contribution to the bispectrum

\begin{equation}
    \Delta B_{\kappa\mathrm{Ly}\alpha}^\mathrm{DLA}(k_\parallel) = b_\mathrm{DLA}\Delta P_{\mathrm{Ly}\alpha}^\mathrm{DLA}(k_\parallel) \sigma^2_{\kappa\bar{\delta}}.
\end{equation}
We chose to set the value $b_\mathrm{DLA}=2.17$, following Ref.~\cite{chiang18}.

%%%%%%%%%%%%%%%%%%%%%%%%%%%%%%%%%%%%%%%%%%%%%%%%%%%%%%%
\section{Estimating the covariance}\label{app:covariance_matrix}

%%%%%%%%%%%%%%%%%%%%%%%%%%%%%%%%%%%%%%%%%%%%%%%%%%%%%%%
\subsection{Estimating the variance of the cross-correlation \& reducing it by subtracting the means}

Here, we explain why subtracting the mean $\kappa$ and $P$ reduces the variance on the correlation $\langle \kappa P \rangle$.
Indeed, for any random variables $\kappa$ and $P$, the variance of the correlator is given by
\beq
\bal
\text{Var}\left[ \kappa P \right]
&=
\langle \left( \kappa P \right)^2 \rangle
- \langle \kappa P \rangle^2 \\
&= 
\langle \kappa^2 \rangle \langle P^2 \rangle
- \langle \kappa \rangle^2  \langle P \rangle^2
\quad\text{if }\kappa \indep P \\
&=
\left( \bar{\kappa}^2 + \sigma_X^2 \right)
\left( \bar{P}^2 + \sigma_P^2 \right)
- \bar{\kappa}^2 \bar{P}^2\\
&=
\sigma_\kappa^2 \sigma_P^2
+
\underbrace{\bar{\kappa}^2 \sigma_P^2 + \bar{P}^2 \sigma_\kappa^2}_\text{additional variance from means}.
\eal
\eeq

In the noise dominated regime, the assumption that the lensing convergence field $\kappa$ is independent from the one-dimensional power spectrum along the line of sight is a good approximation.

%%%%%%%%%%%%%%%%%%%%%%%%%%%%%%%%%%%%%%%%%%%%%%%%%%%%%%%
\subsection{Effect of lensing correlation between neighboring galaxies}\label{app:lensing_patches}

Our analysis correlates the measured $\kappa$ value at the positions of forests with the 1D power spectrum of the same \lya forests.
However, the Wiener-filtered convergence $\kappa$ has a non-zero correlation length, such that lines of sight roughly within a correlation length will have correlated values of $\kappa$.
We can think of splitting the Wiener-filtered convergence map into patches of independent $\kappa$.
In the regime where each patch of independent $\kappa$ contains many forests, we may worry that our uncertainty on the correlation $\langle \kappa P \rangle$ would scale as $1/\sqrt{N_\text{patches}}$ instead of $1/\sqrt{N_\text{galaxies}}$, contrary to our assumption.\\

In this appendix, we explain that this is not the case, as long as the average 1D power spectrum $\bar{P}$ has been subtracted from the measured $P_i$ of each forest $i$.\\

Indeed, our estimator $\hat{C}$ for the correlation between $\kappa$ and $P$ can be expressed as
\beq
\hat{C}
\equiv
\frac{1}{N_\text{galaxies}}
\sum_{i=1}^{N_\text{galaxies}} \kappa_i P_i.
\eeq
We now consider a toy model where the convergence field is constant within patches whose sizes are given by the correlation length of the Wiener-filtered convergence.
We define $N_\alpha$ the number of galaxies in each patch $\alpha=1, ..., N_\text{patches}$, such that 
$\sum_{\alpha=1}^{N_\text{patches}} = N_\text{galaxies}$.
Our estimator can thus be rewritten as
\beq
\hat{C}
=
\frac{1}{N_\text{galaxies}}
\sum_{\alpha=1}^{N_\text{patches}} \kappa_\alpha
\left( \sum_{i=1}^{N_\alpha} P_i \right).
\eeq
The convergence is constant with value $\kappa_\alpha$ within the patch $\alpha$, and independent across patches: $\kappa_\alpha \indep \kappa_{\alpha^\prime}$ for $\alpha \neq \alpha^\prime$.
We further assume that the noise on the 1D power spectrum is independent across galaxies, $P_i \indep P_j$ for $i\neq j$.
Finally, we assume that the noise on $P_i$ and $\kappa_\alpha$ are independent for all $i$ and $\alpha$.
With these assumptions, we can compute the variance of the correlation estimator $\hat{C}$:
\beq
\bal
\text{Var}\left( \hat{C} \right)
&=
\frac{1}{N_\text{galaxies}^2}
\sum_{\alpha=1}^{N_\text{patches}} 
\text{Var} \left[
\kappa_\alpha
\left( \sum_{i=1}^{N_\alpha} P_i \right)
\right]\\
&=
\frac{1}{N_\text{galaxies}^2}
\sum_{\alpha=1}^{N_\text{patches}}
\left[
\langle \kappa_\alpha^2 \rangle
\langle \left( \sum_{i=1}^{N_\alpha} P_i \right)^2 \rangle
-
\langle \kappa_\alpha \rangle^2
\langle \sum_{i=1}^{N_\alpha} P_i \rangle^2
\right]\\
&=
\frac{1}{N_\text{galaxies}^2}
\sum_{\alpha=1}^{N_\text{patches}}
\left\{
\left[ \bar{\kappa}^2 + \sigma_\kappa^2  \right]
\left[ \left( N_\alpha \bar{P} \right)^2 + N_\alpha \sigma_P^2 \right]
-
\bar{\kappa}
\left( N_\alpha \bar{P} \right)^2
\right\}\\
&=
\frac{1}{N_\text{galaxies}^2}
\sum_{\alpha=1}^{N_\text{patches}}
\left\{
\sigma_\kappa^2 N_\alpha^2 \bar{P}^2
+
\left[ \bar{\kappa}^2 + \sigma_\kappa^2 \right] N_\alpha \sigma_P^2
\right\}\\
&= 
\frac{1}{N_\text{galaxies}}
\left( \bar{\kappa}^2 + \sigma_\kappa^2 \right) \sigma_P^2
+
\underbrace{
\frac{1}{N_\text{galaxies}^2}
\left( \sum_{\alpha=1}^{N_\text{patches}} N_\alpha^2 \right)
\sigma_\kappa^2 \bar{P}^2.
}_{\text{only term dependent on }N_\text{patches}\text{ and }N_\alpha}
\eal
\eeq
From the last equation, we can see that the variance has an additional term, function of $N_\text{patches}$ and $N_\alpha$, only if $\bar{P} \neq 0$.
Thus, by subtracting the average power spectrum to the value measured along each forest, we avoid this term completely.
This way, the variance indeed scales as $1/N_\text{galaxies}$.
In particular, when subtracting both the average $\bar{P}$ and $\bar{\kappa}$, the expression above simplifies to the intuitive expectation:
\beq
\text{Var}\left( \hat{C} \right)
=
\frac{1}{N_\text{galaxies}}
\sigma_\kappa^2 \sigma_P^2
\quad\text{if }\bar{\kappa}=0\text{ and }\bar{P}=0.
\eeq

If, on the other hand, we do not subtract the mean convergence and the mean power spectrum, and in the simplifying case where each patch $\alpha$ contains roughly the same number of galaxies $N_\alpha \simeq N_\text{galaxies}/N_\text{patches}$, we get
\beq
\text{Var}\left( \hat{C} \right)
=
\frac{1}{N_\text{galaxies}}
\left(\bar{\kappa}^2 + \sigma_\kappa^2\right) \sigma_P^2
+
\frac{1}{N_\text{patches}}
\sigma_\kappa^2
\bar{P}^2,
\eeq
where the extra term indeed scales as $1/N_\text{patches}$.\\

In summary, by subtracting the average $\bar{P}$ and $\bar{\kappa}$, we ensure that the covariance scales as $1/N_\text{galaxies}$ rather than $1/N_\text{patches}$.
In other words, even though the $\kappa$ values are correlated across lines of sights, the fact that the $P$ values are not means that we can consider the various lines of sight as independent.

%%%%%%%%%%%%%%%%%%%%%%%%%%%%%%%%%%%%%%%%%%%%%%%%%%%%%%%
\subsection{High correlation of the measured data points}
\label{app:correlations_snr}

\begin{table*}[hbp!]
\begin{ruledtabular}
\begin{tabular}{lccc}
 &\multicolumn{3}{c}{\textbf{SNR}}\\
\textbf{Dataset} & No correlation & 30\% constant correlation & Custom correlation (Doux et al.) \\
\hline
Planck+BOSS & 7.8 & 4.7 & 4.4 \\
ACT+DESI-Y1 & 10.8 & 5.9 & 5.7\\
ACT+DESI-Y5 & 19.8 & 10.7 & 10.3\\
SO+DESI-Y5 & 28.4 & 15.4 & 14.8\\
CMB-S4+DESI-Y5 & 39.0 & 21.1 & 20.2
\end{tabular}
\end{ruledtabular}
\caption{Signal-to-noise ratios of the inverse variance weighted signal on the full redshift range. We display the SNR in 3 different cases : without any correlation between $k$ bins, with a constant correlation (30\%) and with a custom correlation inspired from the correlation matrix shown in Ref.~\cite{doux16}}
\end{table*}

%%%%%%%%%%%%%%%%%%%%%%%%%%%%%%%%%%%%%%%%%%%%%%%%%%%%%%%
\section{Forecasts for ACT/CMB-S4 and DESI}
\label{app:forecasts_other_experiments}

\begin{figure}[H]
\centering
\includegraphics[height=0.9\textheight]{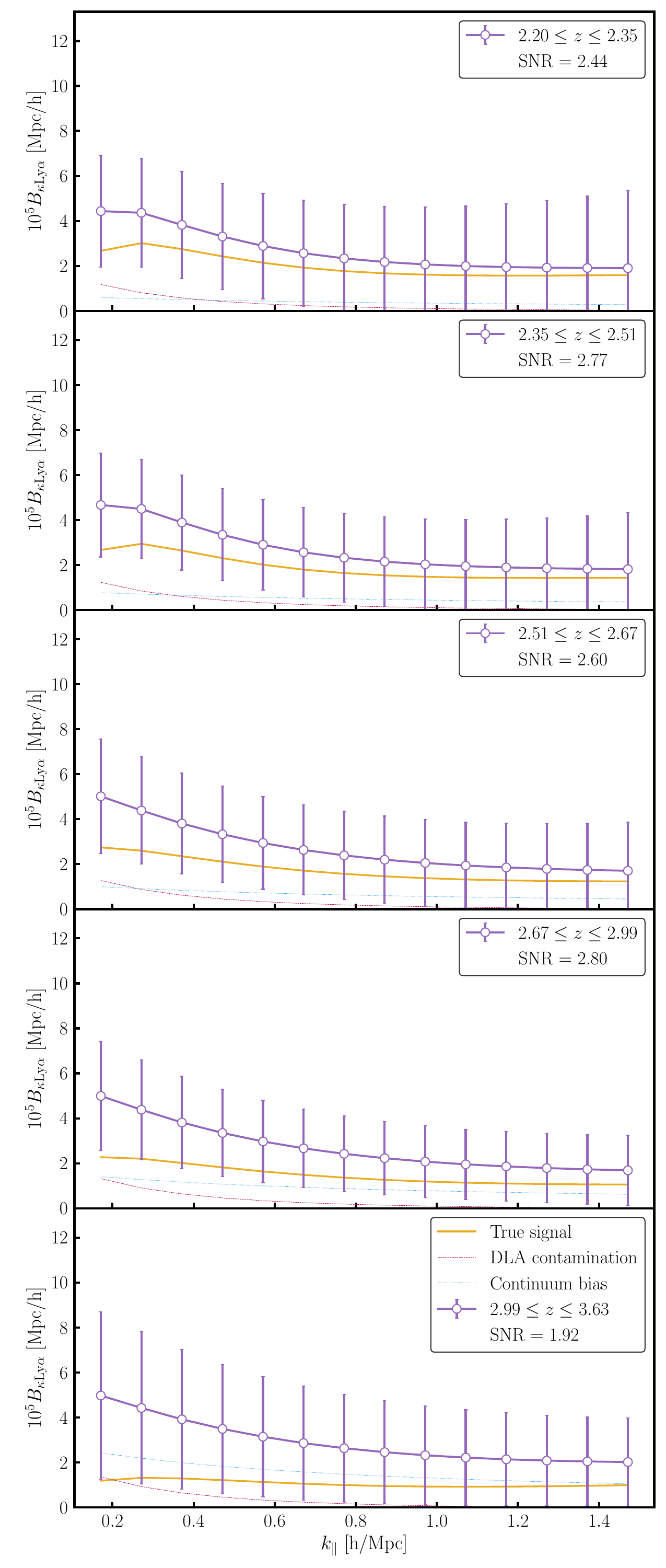}
\caption{Forecasted $\kappa$-$\mathrm{Ly}\alpha$ bispectrum from ACT DR6 lensing and DESI-Y1 $\mathrm{Ly}\alpha$ forest.}
\label{fig:bispectrum_forecast_act_desiy1}
\end{figure}

\begin{figure}[H]
\centering
\includegraphics[height=0.9\textheight]{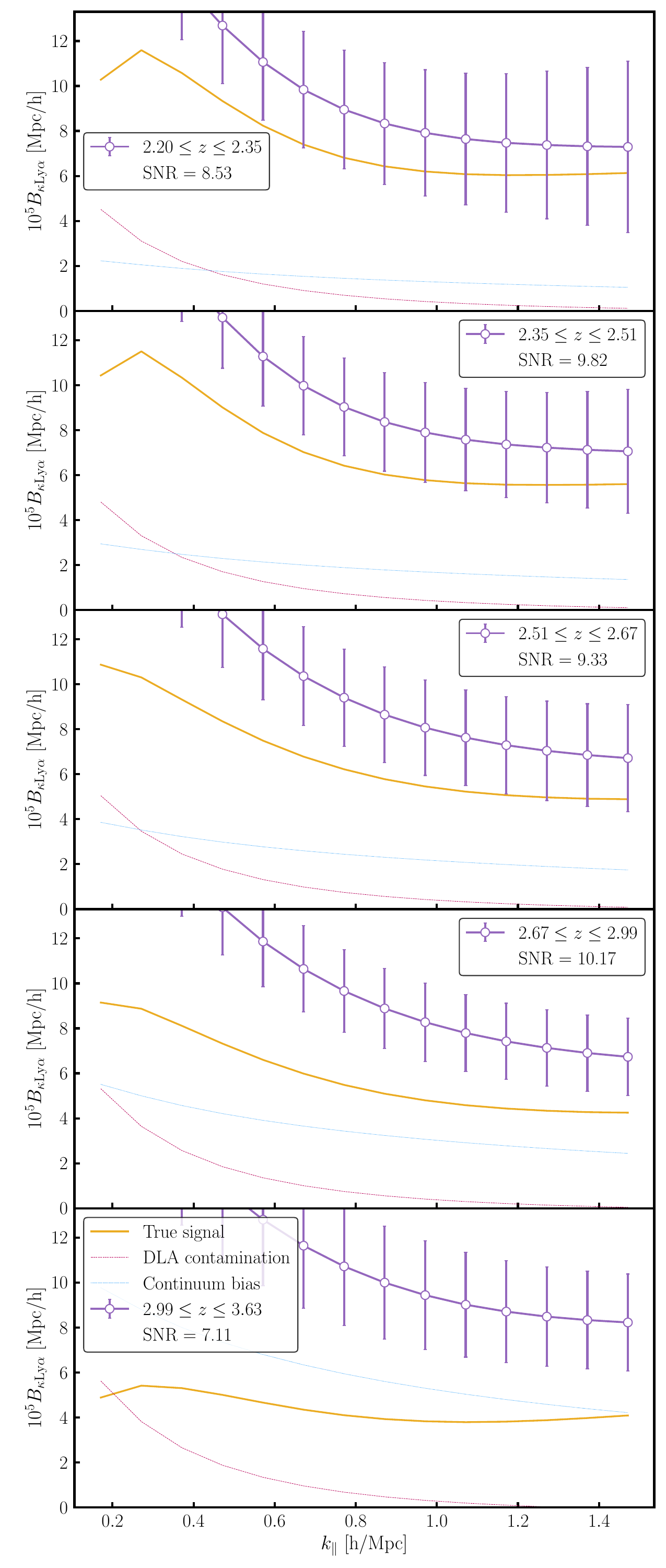}
\caption{Forecasted $\kappa$-$\mathrm{Ly}\alpha$ bispectrum from CMB-S4 lensing and DESI-Y5 $\mathrm{Ly}\alpha$ forest.}
\label{fig:bispectrum_forecast_cmbs4_desiy5}
\end{figure}

%%%%%%%%%%%%%%%%%%%%%%%%%%%%%%%%%%%%%%%%%%%%%%%%%%%%%%%

\end{document}